\documentclass{edp-jp4}
\usepackage{subfigure}
\usepackage[dvips]{graphicx}% Include figure files
\usepackage{dcolumn}% Align table columns on decimal point
\usepackage{bm}% bold math
\newcommand{\eq}{\begin{equation}}
\newcommand{\eqq}{\end{equation}}
\begin{document}
\title{Shape instabilities in vesicles: a phase-field model}
\author{F. Campelo}
\address{Departament d'Estructura i Constituents de la Mat\`{e}ria,\\
Facultat de F\'{\i}sica, Universitat de Barcelona.\\ Diagonal 647, E-08028 Barcelona, Spain}
%\email{campelo@ecm.ub.es}

\author{A. Hern\'{a}ndez-Machado}
\sameaddress{1}
\maketitle
\abstract{
A phase field model for dealing with shape instabilities in fluid membrane vesicles is presented. This model takes into account the Canham-Helfrich bending energy with spontaneous curvature. A dynamic equation for the phase-field is also derived. With this model it is possible to see the vesicle shape deformation dynamically, when some external agent instabilizes the membrane, for instance, inducing an inhomogeneous spontaneous curvature. The numerical scheme used is detailed and some stationary shapes are shown together with a shape diagram for vesicles of spherical topology and no spontaneous curvature, in agreement with known results.
} %end of abstract

\section{\label{sec:introduction}Introduction}

The study of biological membranes has attracted many people from different scientific fields (see \cite{edidin03} for a historical review on cell-membrane models). Physics is one of these fields, nowadays applying its theoretical methods, in addition to the more usual experimental techniques already used many years ago \cite{leibler,safran94,lipowskysackmann}.

Maybe the most important membrane in biology is the plasma membrane, the frontier which defines a cell and separates it in an inside and an outside. This is a very thin wall, usually of the order of a few nanometers, orders of magnitude lower than a typical cell size (a few microns). However, its functionality is much broader than serving as a frontier \cite{alberts}. The high selective permeability of biomembranes is a key point, for instance, in cellular traffic; and the creation of electric potentials in membranes (due to the existence of ion channels and pumps) needed for metabolic regulation, as ATP-formation in mitochondria or signal transduction in neurons. In addition, membranes can be found not only enclosing cells, but also in most of the eukariotic cell organelles. Membranes are composed of several kinds of lipids, which are self-assembled in a fluid bilayer \cite{israelachvili}; and by membrane proteins which are anchored on it \cite{alberts,kozlov06}. From the molecular point of view, biomembranes are extremely complex. However, there seems to be a universal construction principle common to all actual membranes, which is the presence of a fluid lipid bilayer through which proteins can diffuse. Vesicles are closed membranes consisting of one or several different kinds of lipids \cite{seifert97,safran99}. They have therefore been studied to get an idea of the main physical properties of actual biomembranes \cite{lipowskysackmann}.

Since the seminal works by Canham~\cite{canham70} and Helfrich~\cite{helfrich73}, the study of stationary vesicle shapes has been matter of intense research (cfr.~\cite{seifert97} for a review). Many techniques have been used in order to find such shapes in different circumstances. For instance, numerically solving an Euler-Lagrange equation \cite{seifert97}, energy minimization \cite{brakke} or using a phase-field model \cite{campelo06}, among others.

Recently, several experimental results have been reported on dynamic instabilities in membranes, such as pearling~\cite{tsafrir01,tsafrirphd}, budding and tubulation~\cite{tsafrir03,tsafrirphd}. In these experiments shape instabilities are induced by the insertion of a certain concentration (locally or globally) of an amphiphilic polymer (which mimics the proteins within the biomembranes) in the outer leaflet of the bilayer \cite{ringsdorf88}.

The derivation of a dynamic model to study such dynamic instabilities is the aim of this article. More specifically, a dynamic equation for a phase-field which defines the membrane shape is worked out from a free energy functional involving the Canham-Helfrich Hamiltonian with an inhomogeneous spontaneous curvature. Since the effect of the anchorage of amphiphilic polymers on membranes is believed to locally induce a spontaneous curvature on the membrane \cite{tsafrirphd}, our dynamic model would be useful in dealing with those problems.

Phase-field models (or diffuse interface models) can be thought as mathematical tools to study complex interfacial problems, such as free boundary problems \cite{langer86}. Phase-field models are mesoscopic models of the Ginzburg-Landau type, which disregard microscopic details. Such models have been widely used before in different interfacial problems such as solidification and the Saffman-Taylor problem \cite{gonzalezcinca04} and roughening \cite{alava04}. Most of these phase-field models describe the effect of surface tension, but do not deal with bending energies.
 
Our approach considers a conserved dynamic equation which naturally keeps the inner volume of the vesicle constant throughout all the dynamic evolution. Therefore, just one local Lagrange multiplier is needed in order to deal with the incompressibility of the membrane.
 
In this paper we derive a phase-field model for the bending energy of fluid vesicles with an inhomogeneous spontaneous curvature, as in \cite{campelo06}. The membrane is considered as a mathematical interface between two \textit{phases}, the inner fluid and the outer fluid. In this kind of models there is no need to track the interface during the dynamic evolution, which is one of the main problems in membrane dynamics \cite{kraus96}. Our equations are continuous in the whole domain, and the interface is located by the level-set of the phase-field, i.e. the region of rapid variation of the phase-field. The free energy functional associated with this model reduces to the Canham-Helfrich bending energy of the lipid bilayer \cite{{canham70},{helfrich73}} in the so-called sharp interface limit, when interface width goes to zero. In addition, phase-field models are dynamic models, so we are capable with our model to study dynamic properties of vesicles, such as relaxation towards stationary shapes. The fact that we find the correct stationary shapes shows that our free energy functional deals correctly with bending energies.

The organization of this paper is as follows. In section 2 a phase-field model for dealing with the bending energy of fluid lipid bilayers with spontaneous curvature is derived, together with a dynamic equation for the phase-field. The numerical procedure to integrate this dynamic equation is explained in section 3. The results found for this model, and some discussions on that, are presented in section 4. Finally, main conclusions are found in section 5.

\section{\label{sec:model}Model}

\subsection{\label{sec:canham-helfrich}Canham--Helfrich Hamiltonian}

We have just mentioned that a lipid bilayer can be considered a two-dimensional fluid surface embedded in a three-dimensional space. It is thus sensible to mathematically describe this surface, in terms of differential geometry \cite{docarmo76}. A well-behaved two-dimensional surface can be univocally defined given the two radii of curvature at each point or, in other words, its curvature tensor (see Appendix~\ref{surfacegeometry}). Canham~\cite{canham70} and Helfrich~\cite{helfrich73} proposed a Hamiltonian in terms of these curvatures to deal with the energy of a fluid lipid bilayer

\begin{equation}
\mathcal{H}_{\,\mathrm{C-H}}=\frac{\kappa}{2}\int_{\Gamma}{\left(2 H\right)^2+\kappa_G K}\mathrm{d}\mathbf{s},
\label{minimal_model}
\end{equation}
where $\kappa$ and $\kappa_\mathrm{G}$ are two elastic constants: the bending rigidity, and the Gaussian bending rigidity, respectively; $H$ and $K$ are the mean and Gaussian curvatures (see Appendix \ref{surfacegeometry}), respectively.
Due to the Gauss--Bonnet theorem, the Gaussian curvature term (the last term in eq.~(\ref{minimal_model})) integrated over a closed surface is a topological invariant. Since we are not concerned with studying topological changes here, this term will be a constant factor in the total free energy, so it does not need to be considered. Therefore the bending energy reduces to
\eq\label{globalbending}
\mathcal{H}_{\,\mathrm{C-H}}=\frac{\kappa}{2}\int_{\Gamma}\left(2H\right)^2\mathrm{d}\bm{s},
\eqq
where $\Gamma$ is the membrane surface. This model is the simplest possible model for lipid bilayers. There are other models which include further terms (cfr. \cite{seifert97,safran99} for reviews). One of this models is the so-called spontaneous curvature model, which takes into account a possible asymmetry between the tow leaflets of the bilayer. This asymmetry induces a preferential curvature to the bilayer, $c_0$, the spontaneous curvature. The corresponding Hamiltonian reads as
\eq\label{globalbending_sc}
\mathcal{H}_{\,\mathrm{C-H},\ \mathrm{sc}}=\frac{\kappa}{2}\int_{\Gamma}\left(2H-c_0\right)^2\mathrm{d}\bm{s}.
\eqq

\subsection{\label{sec:themodel}Phase--field implementation}

A phase-field dependent free energy for the Canham-Helfrich Hamiltonian with spontaneous curvature was derived in Ref.~\cite{campelo06}. This free energy functional is
\eq
\mathcal{F}_{\mathrm{sc}}[\phi]=\int_{\Omega}{\Phi^2_{\mathrm{sc}}[\phi\,]\ \mathrm{d}\bm{x}},
\label{ansatz_sc}
\eqq
where
\eq\label{chem.pot}
\Phi_{\mathrm{sc}}[\phi(\bm{x})]=\left(\phi-\epsilon\,C_0(\bm{x})\right)\,(\phi^2-1)-\epsilon^2\, \mathbf{\nabla}^2 \phi,
\eqq
where $\phi(\mathbf{x})$ is the phase-field, $\epsilon$ is a small parameter related to the interface width, and $C_0(\bm{x}) = c_0(\bm{x})/\sqrt{2}$ is related to the spontaneous curvature which, in principle, can be position-dependent (or even $\phi$-dependent). 

The minimum of the free energy Eq.~(\ref{ansatz_sc}), with no constraints and zero spontaneous curvature, is obtained by setting Eq.~(\ref{chem.pot}) equal to 0. In one dimension, this leads to the tanh-like solution $\phi(x)=\tanh{\left(\frac{x}{\sqrt{2}\epsilon}\right)}$, given the boundary conditions $\phi(\pm \infty)=\pm 1$. The boundary conditions in three dimensions are that the phase-field at infinity is $\phi=-1$, which is the value for the stable phase of the outside bulk.

\subsubsection{\label{sec:geom.constraints}Geometrical constraints}

The shapes of lipidic vesicles may be subject to certain geometrical constraints.

On one hand, at physiological relevant temperatures, lipid membranes are usually in a liquid disordered phase. In addition, the membrane can be considered to be locally incompressible. Besides, the solubility of membrane lipids is extremely low, which implies no relevant exchange of material between the membrane and the surrounding media. Therefore, in the absence of high enough thermal fluctuations, these two facts provide us a constraint for the local area of the vesicle, which remains fixed.

On the other hand, biological membranes are permeable to water, but not to, e.g., large ions (on the time scales we are interested in) \cite{alberts}. This means that any transfer of water through the membrane would create an osmotic pressure which cannot be counterbalanced by the relatively much weaker bending energy,\cite{helfrich73}. Therefore, the concentration of osmotically active molecules fixes the inner volume of the vesicle.

The usual way to implement these conditions in the free energy is introducing a Lagrange multiplier for each constraint. This method has shown to be very useful in finding the stationary vesicle shapes \cite{seifert97}. The Lagrange multiplier coupled with the surface area being interpreted as a surface tension, and the one coupled with the inner volume, as an osmotic pressure. Therefore, if one wants to include a surface tension or an osmotic pressure instead of keeping the surface area or the inner volume constant, Lagrange multipliers would just be those physical quantities.

The implementation into the phase-field model of these geometrical constraints can be achieved be extending the free energy functional eq.~(\ref{ansatz_sc}). An effective free energy functional with two Lagrange multipliers can be thus written. We choose \cite{campelo06} to use this formalism in order to include a surface term, since it is sometimes useful to be able to switch between the Lagrange multiplier and the surface tension points of view. The effective free energy functional is thus defined
\eq
\mathcal{F}_{\mathrm{eff}}[\phi]=\mathcal{F}_{\mathrm{sc}}[\phi]+\int_{\Omega}{\sigma(\bm{x}) a[\phi]\mathrm{d}\bm{x}},
\label{effective0}
\eqq
where $\mathcal{F}[\phi]$ is given by eq.~(\ref{ansatz_sc}), $\sigma$ is the local Lagrange multiplier, and $a[\phi]$ is the local surface area functional,
\eq
a[\phi]=\frac{3}{2\sqrt{2}}\ \epsilon\left|\bm{\nabla} \phi\right|^2,
\eqq
which is the phase-field representation of the area density, such that $\int_{\Omega}{a[\phi]\mathrm{d}\bm{x}}=\int_{\Gamma}\mathrm{d}\bm{s}$, where $\Gamma$ is the two-dimensional vesicle surface embedded in the three-dimensional domain $\Omega$.

Volume conservation is implemented dynamically by using a model-B like dynamics, namely
\eq\label{cons.dyn}
\frac{\partial \phi}{\partial t}=\bm{\nabla}^2\left(\frac{\delta\mathcal{F}_{\mathrm{eff}}}{\delta\phi}\right).
\eqq
This dynamic equation ensures that $\int_{\Omega}{\phi(\bm{x})\mathrm{d}\bm{x}}$ is constant in time.

\subsection{\label{sec:dyn.eqn}Dynamic equation}

The dynamic relaxation towards free energy minima is achieved in our model by conserved relaxation dynamics, eq.~(\ref{cons.dyn}). Relaxational dynamics \cite{foltin94} have been used before, for instance, to study phase-separation dynamics of two-component vesicles \cite{taniguchi96}. In our phase-field approach, we need to compute the functional derivative in eq.~(\ref{cons.dyn}). This calculation leads to the following dynamic equation for the phase-field $\phi(\mathbf{x})$,
% \begin{eqnarray}\label{dyn.eqn}
\eq\label{dyn.eqn}
\frac{\partial \phi}{\partial t}=2\bm{\nabla}^2\Big\{\left(3\phi^2-1+2\epsilon C_0(\bm{x})\, \phi\right)\Phi_{\mathrm{sc}}[\phi]
-\epsilon^2\bm{\nabla}^2\Phi_{\mathrm{sc}}[\phi]+\epsilon^2\bar{\sigma}(\bm{x})\bm{\nabla}^2 \phi\Big\},
\eqq
% \end{eqnarray}
where $\bar{\sigma}$ is defined as
\eq\label{lagr.mult}
\bar{\sigma}(\bm{x})=\frac{\sqrt{2}}{3 \epsilon}\sigma(\bm{x}).
\eqq

Using this kind of dynamics, local conservation of the inner volume of the vesicle is achieved in a natural way, unlike Ref.~\cite{duliuwang06} which uses a purely relaxational dynamics with no direct conservation of the inner volume of the vesicle.

In addition, a term proportional to $\bm{\nabla}\bar{\sigma}(\bm{x})$ in the dynamic equation (\ref{dyn.eqn}) could be wrote down. However, it was shown to be small \cite{campelo06}, so the Lagrange multiplier, $\bar{\sigma}$, can be considered homogeneous. Moreover, $\sigma(\bm{x})$ appears as an effective surface tension which prevents the surface area from changing. Anyway, its value in membranes is very small compared with other energy scales in the system (e.g. bending rigidity) \cite{evans90}. Therefore, its variations are also small.

\section{\label{sec:num.int}Numerical Integration}

As we have argued in the introduction of this paper, phase-field models are methods for dealing with moving boundary problems by means of solving partial differential equations for some order parameters. Usually, these partial differential equations are highly non-linear, and a numerical procedure to solve them is needed. Our dynamic equation is not an exception, notice, e.g., the coupling between the field $\phi^2$ and the functional $\Phi[\phi]$.

The discretization algorithms used are second-order finite differences for the spatial dependence, and an Euler scheme for the time dependence \cite{strikwerda}. Since standard second-order finite differences is a consistent finite difference method, the time step was chosen following the Courant-Friedrichs-Lewy stability criterion, $\Delta t\le |k|\ \Delta x$, where $k$ is some constant. We can thus assume that the algorithms used are convergent \cite{bertsekas}. 

Our effective free energy functional (\ref{effective0}) explicitly contains a Lagrange multiplier. Therefore, we need to know the time evolution of the Lagrange multiplier. To do this, we have used a first order Lagrangian method \cite{bertsekas}. Lagrangian methods can be formally written as
\eq
\left.
\begin{array}{l}
\phi^{k+1}(\bm{x})=G(\phi^k(\bm{x}),\sigma^k(\bm{x}))\\
\sigma^{k+1}(\bm{x})=H(\phi^k(\bm{x}),\sigma^k(\bm{x}))
\end{array}
\right\},
\eqq
such that,
\begin{eqnarray}
\phi^*(\bm{x})=G(\phi^*(\bm{x}),\sigma^*(\bm{x})),\nonumber\\
\sigma^*(\bm{x})=H(\phi^*(\bm{x}),\sigma^*(\bm{x})),
\end{eqnarray}
are the stationary values. The simplest of these methods is the first-order method, given by
\eq
\sigma^{k+1}(\bm{x})=\sigma^k(\bm{x})+\alpha \left(a[\phi^k(\bm{x})]-a_0(\bm{x})\right),
\eqq
together with the dynamic equation for the phase-field (\ref{dyn.eqn}). $\alpha>0$ is the stepsize, and $a_0(\bm{x})$ is the fixed local surface area. Since we are not interested in the actual dynamics of the multiplier, our choice is justified because it does not change the dynamics of the phase-field, but it just keeps the surface area of the vesicle constant during the time evolution without altering the dynamics.

In order to study the dynamics of a shape evolution or of a shape instability, we need to prepare the system with the desired initial shape. The initial shape corresponds to the initial values of the phase-field. According to the diffuse-interface nature of phase-field models, we choose to consider as initial shape a shape with an already created diffuse interface. In order to get such an interface, we start with a sharp interface and let it evolve under the dynamic equation (\ref{dyn.eqn}) under no constraints. After some time steps, the shape hardly changes, but a tanh-like profile in the interface is rapidly created. This is the initial shape used to compute, for instance, the surface area.  

We have performed simulations on lattices of different sizes and equivalent shapes and evolutions were obtained. In addition, during the time evolution, we checked the value of the free energy evolution in time to see how it relaxes to a stationary value in a monotonically decreasing way. The values of the inner volume and the surface area were also computed during the evolution and it can be seen that the volume remains constant (up to the numerical precision) during all the process, and similarly with the surface area (the value of the Lagrange multiplier converges rapidly to the stationary solution).

\section{Results and Discussion}

In our model there seem to be several free parameters ($\epsilon$, $a_0$, $V_0$, $C_0(\mathbf{x})$). However, $\epsilon$ is a small parameter (the model is shown to be robust under variations of this parameter), which will be set, in what follows, to be equal to the mesh size. In addition, scale invariance causes that the ratio between the constrained total volume and the total surface area is the only relevant parameter in the model (for a fixed topology). Thus, we define a dimensionless volume $v$ as the ratio between the actual volume and the volume of a sphere with the same area, $v=\frac{V}{(4\pi/3) R_0^3}$, where $R_0=\left(\frac{A}{4 \pi}\right)^{1/2}$. The function $C_0(\mathbf{x})$ is the spontaneous curvature, and is given by the intrinsic properties of the lipid bilayer and by the local insertion of curvature-inducing anchor groups into the bilayer.

\begin{figure}[htbp!]
\centering
\vskip5mm
\includegraphics[angle=0,width=11.5cm]{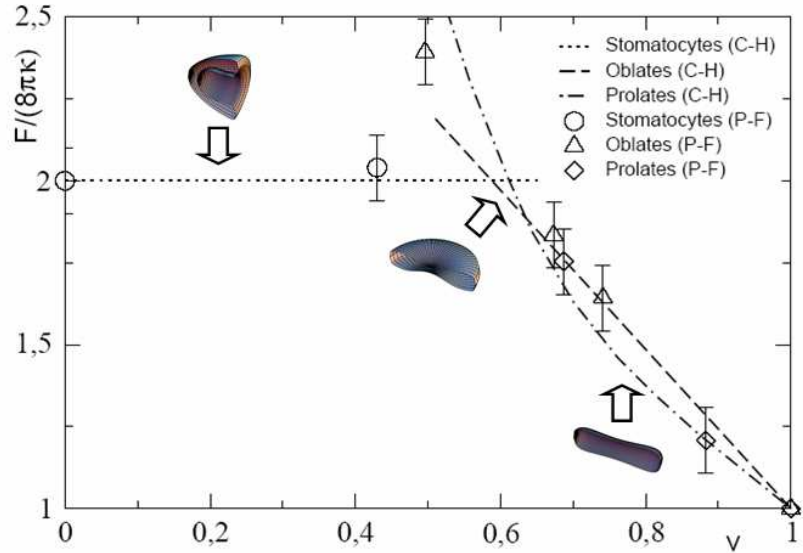}
\caption{Shape diagram for vesicles of spherical topology corresponding with a model with no spontaneous curvature. Lines correspond to standard Euler-Lagrange minimization of the Canham-Helfrich free energy \cite{seifert91pra}. Symbols are the results found using a long-time relaxation of our dynamic phase-field model. The three different kinds of shapes, stomatocytes, oblates and prolates, are also shown, respectively, from left to right.}
\label{fig:shapediagram}
\end{figure}

The validity of the bending phase-field model with all the geometric constraints is checked by its stationary shapes. The shape diagram of closed vesicles with spherical topology and vanishing spontaneous curvature has been worked out from our phase-field simulations and compared (see fig.~\ref{fig:shapediagram}) with already known results from purely equilibrium techniques \cite{seifert97}.

\section{Conclusions}

In order to study dynamic instabilities in membranes, in this paper we have derived a dynamic equation for a phase-field model for the bending energy of bilayers with an inhomogeneous spontaneous curvature. Knowing this field in each point, means knowing the shape of the vesicle at each time step. By letting the shape relax, it is possible to find stationary shapes of vesicles for different spontaneous curvatures (both homogeneous, inhomogeneous or even non-constant in time). 

Besides, we checked the numerics linked to this model, and we saw the convergence of the model. Free energy relaxation is seen, as well as convergence of the Lagrange methods. The numerical algorithms used to solve the partial differential equations are also seen to be stable, and robustness of the numerical parameters is preserved.

Finally, stationary shapes for vesicles with no spontaneous curvature and spherical topology are presented here, together with a shape diagram which is in good agreemet with already known equilibrium results.

\begin{acknowledgements}

We are grateful to Joel Stavans for drawing our attention to the problem of membranes. We acknowledge financial support of the Direcci\'{o}n General de Investigaci\'{o}n under project No. BFM2003-07749-C05-04. F.C. also thanks Ministerio de Educaci\'{o}n y Ciencia (Spain) for financial support.

\end{acknowledgements}

\appendix

\section{Surface Geometry}\label{surfacegeometry}

% 
% Being $\bm{R}(\sigma_1,\sigma_2)$ the position vector defining the surface parametrized by $\sigma_1$ and $\sigma_2$, its tangent vectors are defined as
% \eq
% \bm{R_i}= \partial_i \bm{R}.
% \eqq
% A unit vector normal to the surface is therefore defined as
% \eq
% \hat{n}=\frac{\bm{R}_1 \times \bm{R}_2}{\left|\bm{R}_1 \times \bm{R}_2\right|}.
% \eqq
% The curvature tensor is 
% \eq\label{curv.tensor}
% h_{ij}= \left(\partial_i \partial_j \bm{R}\right)\cdot \hat{n}.
% \eqq
% The two invariants of this tensor \cite{docarmo76} are related with the mean and the Gaussian curvature by
% \begin{eqnarray}\label{tns.inv}
% H&=& -\frac{1}{2}\mathrm{tr}\, h^i_j \nonumber \\
% K&=& \mathrm{det}\left(h_j^i\right),
% \end{eqnarray}
% where the raising of indexes in the curvature tensor is done in the usual way by the metric tensor, defined by
% \eq
% g_{ij}= \bm{R_i}\cdot\bm{R_j}.
% \eqq
% 
% 
% **********************************

From a geometrical point of view, a two-dimensional surface embedded in a three-dimensional space can be represented using the so-called Monge representation \cite{docarmo76,kamien02}, where the z-axis is parametrized by a \emph{height}-coordinate depending on the other two Cartesian coordinates, $x$ and $y$. Namely
\eq\label{monge}
z=h(x,y).
\eqq
In addition, such a surface, supposed to be well-behaved, can always be characterized by two radii of curvature, $r_1$ and $r_2$, since it can be locally approximated by an ellipsoidal surface
\eq\label{radii}
z(x,y)=\sqrt{ \frac{x^2}{r_1(x,y)}+\frac{y^2}{r_2(x,y)}}
\eqq
Two geometrical invariants can be defined in terms of these radii, the mean curvature, $H$, and the Gaussian curvature, $K$:
\begin{eqnarray}\label{invariants}
H&=&\frac{1}{2}\left(\frac{1}{r_1}+\frac{1}{r_2}\right) \nonumber \\
K&=&\frac{1}{r_1 r_2}.
\end{eqnarray}
These are the two invariants of a second-order tensor, the curvature tensor, which is defined as
\eq\label{curv.tensor}
h_{ij}= \left(\partial_i \partial_j \bm{R}\right)\cdot \hat{n},
\eqq
where $\bm{R}$ is the position vector defining the surface, and $\hat{n}$ is a unit vector normal to the surface. Therefore the two invariants of this tensor would be its trace and its determinant. It can be proved \cite{docarmo76} that they are related with the mean and the Gaussian curvature by
\begin{eqnarray}\label{tns.inv}
H&=& -\frac{1}{2}\mathrm{tr}\, h^i_j \nonumber \\
K&=& \mathrm{det}\left(h_j^i\right),
\end{eqnarray}
where the raising of indexes in the curvature tensor is done in the usual way by the metric tensor, defined by
\eq
g_{ij}= \bm{R_i}\cdot\bm{R_j},
\eqq
where $\bm{R_i}$ are the vectors tangent to the surface
\eq
\bm{R_i}= \partial_i \bm{R}.
\eqq
Therefore it follows from (\ref{tns.inv}), that the mean curvature is proportional to the Laplacian of the normal component of the position vector, so it is sensible to assert that:
\eq
H\sim \bm{\nabla} ^2 h(x,y).
\eqq

\bibliographystyle{h-physrev3}

\bibliography{/home/felix/mypapers/bib}

\end{document}